# Beam Dynamical Evolutions in a Solenoid Channel – A Review

[*]H.F. Kisoglu[1,2], M. Yilmaz[2]

[1]*Depertment of Physics, Aksaray University, Aksaray, Turkey*
[2]*Physics Department, Gazi University, Ankara, Turkey*

**Abstract**

Today a linear particle accelerator (linac), in which electric and magnetic fields are of vital importance, is one of the popular energy generation sources like Accelerator Driven System (ADS). A multipurpose, including primarily ADS, proton linac with energy of ~2 GeV is planned to constitute within the Turkish Accelerator Center (TAC) project collaborated by more than 10 Turkish universities. A Low Energy Beam Transport (LEBT) channel with two solenoids is a subcomponent of this linac. This solenoid channel transports the proton beam ejected by a proton source, and matches it with the Radio Frequency Quadrupole (RFQ) that is a subcomponent just after the LEBT. These solenoid magnets are used as focusing element to get the beam divergence and emittance growth under control. This paper includes settings of the LEBT solenoids with regard to beam dynamics, which investigates the beam particles motion in particle accelerators, for TAC proton linac done by using a beam dynamics simulation code PATH MANAGER. Furthermore, the simulation results have been interpreted analytically.

---

[*] Corresponding author. E-mail:hasanfatihk@aksaray.edu.tr

## 1. Introduction

Particle accelerators are generic technologies widely used in many fields of science and technology [1]. These physics devices uses electric and magnetic fields to accelerate and to orient a charged particle or ion beam. Today, there are too many all-duty accelerator complexes around the world. The main ones such as the European Organization for Nuclear Research (CERN) [2], Fermilab [3], Los Alamos National Laboratory (LANL) [4], … serve as a model for others. In this manner, the Turkish Accelerator Center (TAC) project, ratified by Turkish State Planning Organization (DPT) in 2006, aims at being a regional project [5]. It would be a multi-component complex in a way to provide service to many users, and it is brought into being by more than 10 Turkish universities [6].

A linear proton accelerator (proton linac) is planned to give proton beam an acceleration of ~2 GeV so as to serve as a supplier for many areas of utilization from industry to medical [7] within the TAC. It can be used to research in nuclear science and high energy physics. The most important goal of the project is to realize the Accelerator Driven System (ADS) using the thorium reserves of Turkey efficiently [8].

The solenoid channel, named as Low Energy Beam Transport (LEBT), would be placed at the front end of the envisioned proton linac. It will transport and match the proton beam ejected by a proton source to following part, Radio Frequency Quadrupole (RFQ), via magnetic fields provided by two solenoid magnets. These solenoid magnets are used as focusing element to get the beam divergence and emittance growth [9] under control.

In this study, we present the settings of the LEBT solenoids with regard to beam dynamics, which investigates the beam particles motion in particle accelerators, for TAC proton linac. Furthermore, these simulation results which shows the behavior of the beam inside the solenoid have been interpreted according to electromagnetic theory. The specifications of the considered LEBT channel are given in Table 1.

**Table 1.** Specifications of the LEBT channel

| Parameters | Length (mm) |
|---|---|
| **Drift** | 200 |
| **Solenoid** | 300 (600 mm effective) |
| **Drift** | 802 |
| **Solenoid** | 300 (600 mm effective) |
| **Drift** | 350 |
| **Beam aperture** | 55 |

We have used two Linac4 [10] solenoids as the focusing elements, and input beam current of 33 mA has been chosen for beam dynamics of the LEBT channel in accordance with the latest feasibility studies. The beam dynamics has been simulated using PATH MANAGER [11] in this paper.

## 2. Beam Dynamics Based Settings of the LEBT Solenoids

The specifications and schematic layout of the LEBT are given in Table 1 and Figure 2, respectively. Such a LEBT channel will be located between a proton source and an RFQ. This channel will transport and match a proton beam with 30–40 mA current and ~50 keV kinetic energy to RFQ. Two solenoid magnets are planned to use to keep the beam divergence and emittance growth under control.

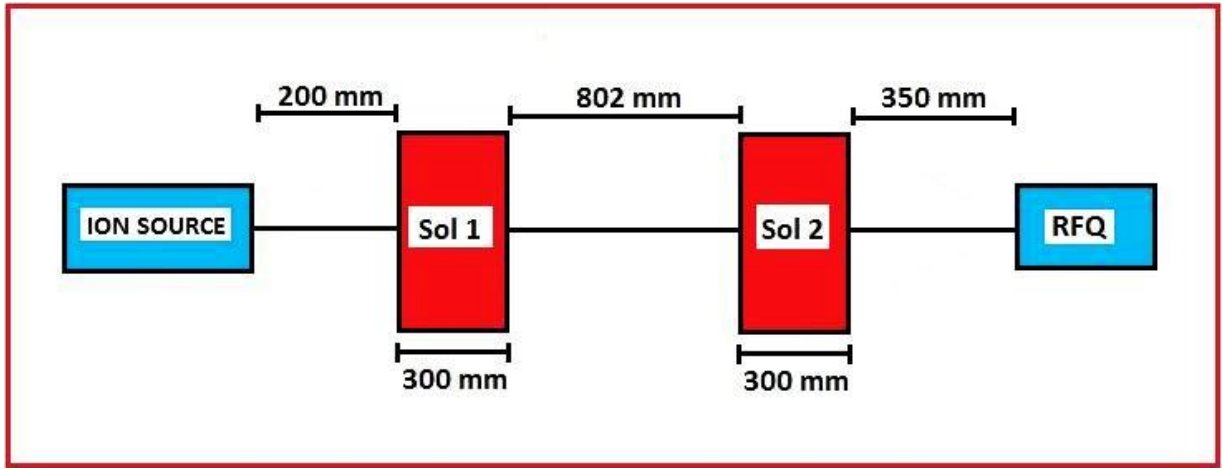

**Figure 2.** Schematic layout of the LEBT channel

In beam dynamics simulation of the LEBT, we have assumed that proton beam is ejected by the source with a current of 33 mA and kinetic energy of 50 keV. Furthermore, the particle distribution of such a beam has been chosen as 4-D uniform distribution with 10000 particles. The figures of merit for the beam dynamics simulation of the LEBT channel were getting the appropriate beam transmission and emittance values to accord with the RFQ [12]. The magnetic fields of the solenoids have been tuned in this manner.

We have assumed the initial emittance at the beginning of LEBT as $0.11\times10^{-6}$ π·m·rad (normalised, rms) on both ($x$, $x'$) and ($y$, $y'$) phase-spaces ($xy$-plane is perpendicular to the $z$ beam axis) for the beam dynamics simulation. The magnetic fields of the solenoids have been tuned using Delta option of PATH MANAGER to acquire the emittance (norm., rms) and transmission values equal to ones of input beam of the RFQ, i.e., 0.20 π·mm·mrad and 30 mA, respectively. Main reason of such a decreasing on the current is repulsive space-charge forces between the beam particles. The transverse $B_r(z)$ and longitidunal $B_z(z)$ magnetic fields of both solenoids which give the appropriate transmission value regarding the RFQ input current, i.e. ~90% (Figure 3), are given in Figure 4. The evolution of the beam emittance and divergence (i.e. momentum) along the LEBT channel is given in Figure 5 and Figure 6, respectively and the results of the beam dynamics simulation of the LEBT channel are given in Table 2.

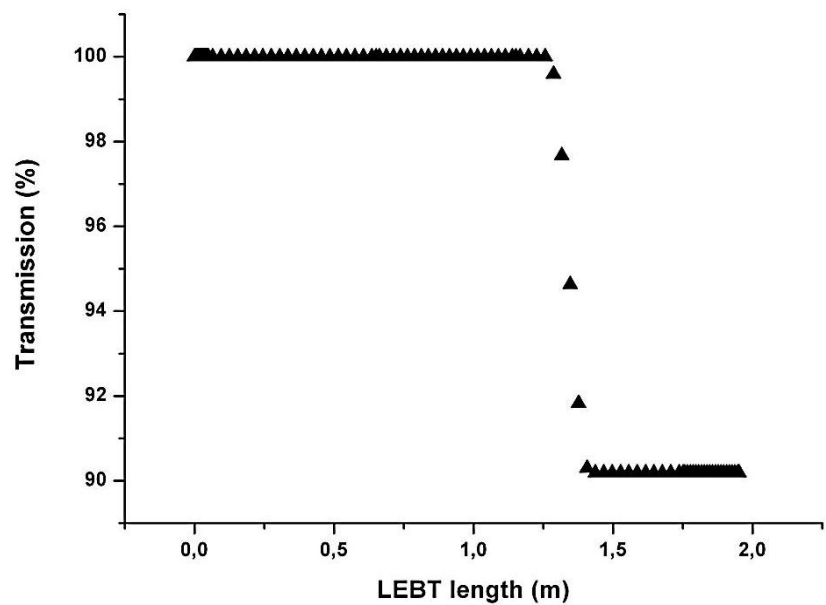

**Figure 3.** Beam transmission along whole LEBT

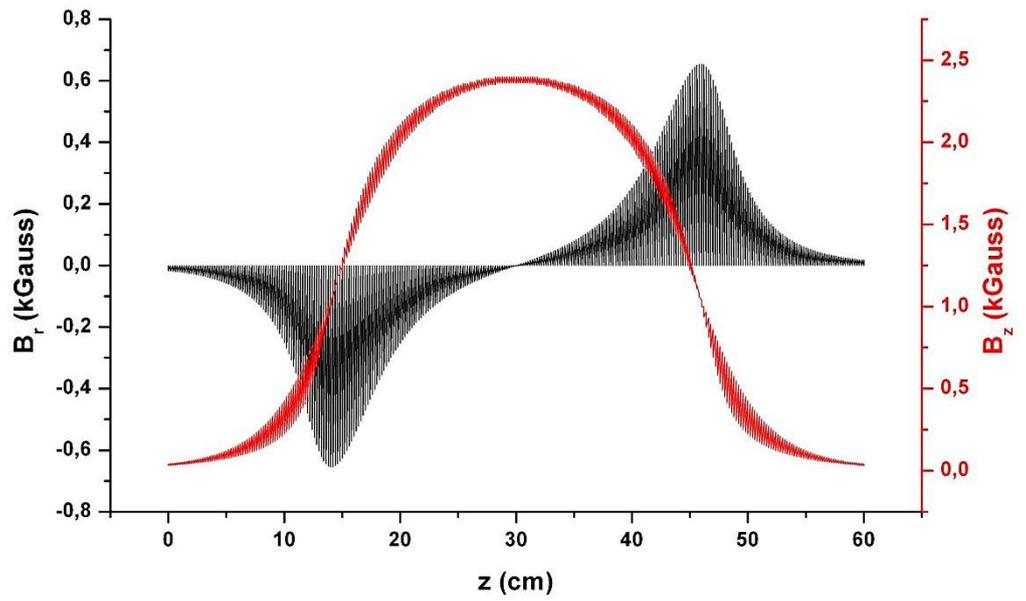

(a)

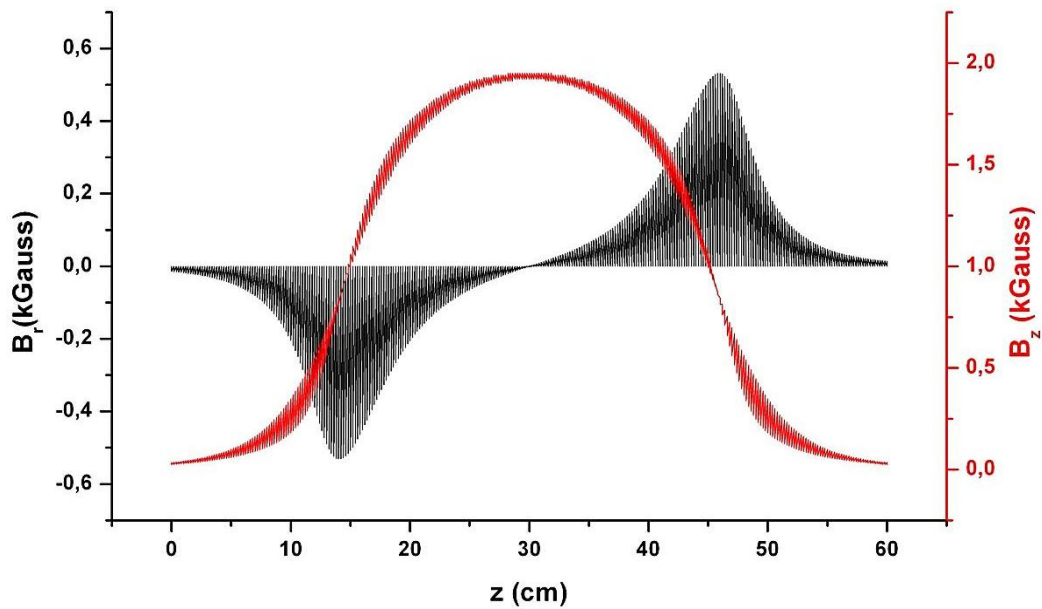

(b)

**Figure 4.** Transverse ($B_r(z)$) and longitudinal ($B_z(z)$) magnetic field patterns of sol.1 (a) and sol.2 (b) which provide minimum emittance growth and approporiate transmission value

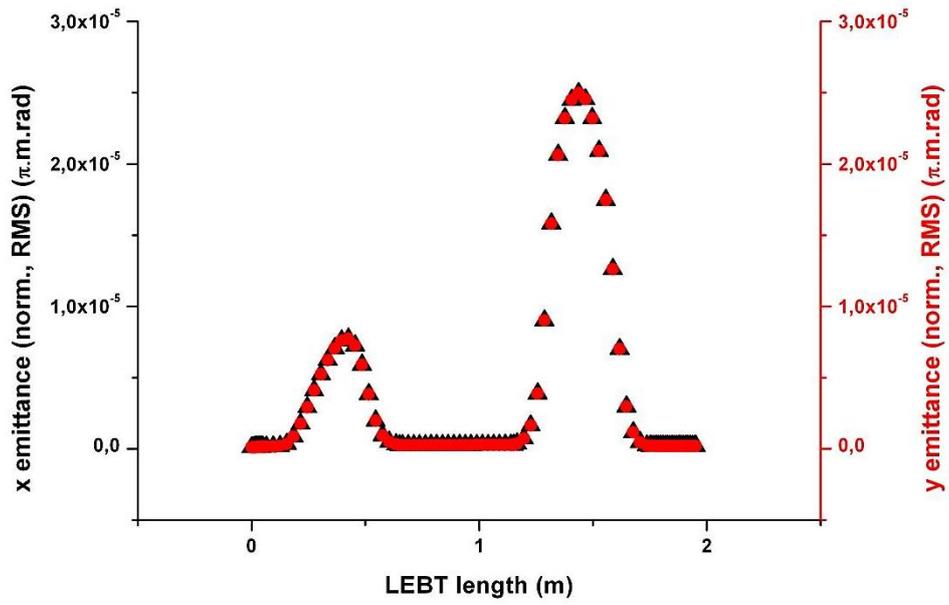

**Figure 5.** Emittance evolution along the LEBT line.

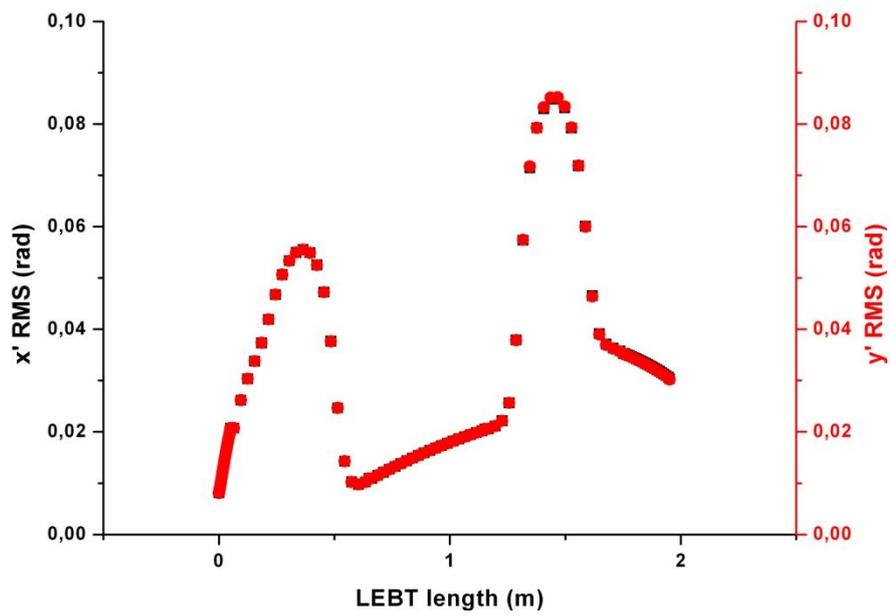

**Figure 6.** Beam divergence (momentum) evolution along the LEBT line

**Table 2.** Beam dynamics simulation results of the LEBT

| Max. magnetic field of sol.1 (kG) | Max. magnetic field of sol.2 (kG) | Norm. rms emittance at the exit ($\pi$·m·rad) | Transmission (%) |
|---|---|---|---|
| 0.65 ($B_r$) | 0.53 ($B_r$) | $0.20 \times 10^{-6}$ ($\varepsilon_x$ and $\varepsilon_y$) | 90.19 |
| 2.40 ($B_z$) | 1.95 ($B_z$) | | |

## 3. Analytical Approach to Emittance Evolution in Solenoids

Solenoids are used to focus the beam on the $z$-beam axis in a LEBT channel. But, as is seen from Figure 5, there are growths on transverse emittance from entry to mid of both solenoids. This, although a predictable outcome, may be confusing.

In order to support the simulation results in the previous section, emittance evolution in the solenoids is discussed analytically in this section. Effetcs that cause the transverse emittance growth have been handled in this approximation for which a charged particle passing through a solenoid.

A general expression for the magnetic field of a solenoid can be obtained from Maxwell's equations in vacuum given as follows;

$$\vec{\nabla} \times \vec{B} = \mu_0 \vec{J} + \mu_0 \varepsilon_0 \frac{\partial \vec{E}}{\partial t} = 0 \tag{1}$$

Equation (1) allows us to write the magnetic field in terms of a scalar potential, $\xi$;

$$\vec{B} = -\vec{\nabla} \xi \tag{2}$$

In this case, we can get the Laplace equation for $\xi$;

$$\nabla^2 \xi = 0 \tag{3}$$

Eq. (3) can be written in cylindrical coordinates in accordance with the geometry of the solenoid;

$$\frac{\partial^2 \xi}{\partial r^2} + \frac{1}{r}\frac{\partial \xi}{\partial r} + \frac{\partial^2 \xi}{\partial z^2} + \frac{1}{r}\frac{\partial^2 \xi}{\partial \varphi^2} = 0 \qquad (4)$$

where $z$ is the axis of the solenoid and $r$ is the radial distance from axis of the solenoid. We can ignore the last term at the left side of this equation due to the axial symmetry of system. Hence, $\xi \equiv \xi(r,z)$, solutions for scalar potential are obtained. Equation (4) can be solved by numerical methods. But we can suppose solutions as following [13];

$$\xi_{(r,z)} = \sum_k b_{k(z)} r^k \qquad (5)$$

where $b_{0(z)} = \xi_{(z)}$ is the solution on the axis of the solenoid. If we replace the potential solutions in Eq.(5) into the Eq.(3) and rearrange the $k$ indexes we get the following conditions;

$$b_{2k} = \frac{(-1)^k}{2^{2k}(k!)^2} \xi_{(z)}^{(2k)} \qquad \forall k \geq 2$$

$$b_{2k+1} = 0 \qquad \forall k \qquad (6)$$

Hereat, the potential expression in Eq.(5) turns into;

$$\xi_{(r,z)} = \sum_k \frac{(-1)^k}{(k!)^2} \left(\frac{r}{2}\right)^{2k} \xi_{(z)}^{(2k)} \qquad (7)$$

If this equation is substituted into Eq.(2), the magnetic field components of the solenoid can be obtained as follows;

$$B_r = \sum_{k=1}^{\infty} \frac{(-1)^k}{k!(k-1)!} \left(\frac{r}{2}\right)^{2k-1} B_{(z)}^{(2k-1)}$$

$$B_\varphi = 0 \tag{8}$$

$$B_z = \sum_{k=0}^{\infty} \frac{(-1)^k}{(k!)^2} \left(\frac{r}{2}\right)^{2k} B_{(z)}^{(2k)}$$

The higher order terms in $B_r$ can be neglected in the linear conditions of the solenoid. As a conclusion, the transverse component of solenoid magnetic field is got as follows;

$$\vec{B}_r = -\frac{r}{2} \frac{\partial \vec{B}_{(z)}}{\partial z} \tag{9}$$

According to this result, a variation of the solenoidal magnetic field on axis of the solenoid, i.e. at the start/end edges, induces a variance on the transverse magnetic field component of the solenoid in the opposite direction. The particle is focused/defocused at the edges as a result of this variance. This outcome agrees with Figure (4). As is seen, the transverse magnetic field component begins to increase negatively as the longitudinal component rises positively.

According to Figure (4-a), $B_z$ increases between $z=$ 0–30 cm and decreases between $z=$30–60 cm. That is why the $B_r$ increases negatively and positively between $z=$ 0–30 cm and $z=$30–60 cm, respectively. Furthermore, $B_z$ increases in a "sawtooth wave" form between $z=$ 0 and ~15 cm. The amplitude of this "wave" is minimum at $z=$ ~15 cm and it has a slow positive ramp. Beside, $B_r$ reaches negatively maximum value at this point. Between $z=$ ~15 cm and 30 cm, the "sawtooth wave" has a slow negative ramp and $B_r$ shapes on the contrary.

At $z=$30 cm, $B_z$ is maximum, hence, $B_{(z)}^{'} = 0$ and $B_r$ =0. Between $z=$30–60 cm, we can say the similar things for the evolution processes belong to $B_z$ and $B_r$. All that we have said for first solenoid (Fig. 4-a) is also true for the second one (Figure 4-b).

As for the magnetic force which is given as;

$$\vec{F} = q\vec{v} \times \vec{B} \tag{10}$$

and experienced by charged particle passing through the $\vec{B}_r$ transverse magnetic field with a velocity of $\vec{v}_z$ is on the azimuthal, i.e., transverse plane and can be formed hereinbelow;

$$F_\varphi = qv_z B_r = \frac{dp_\varphi}{dt} \tag{11}$$

If we use the chain rule, $\frac{dp_\varphi}{dt} = \frac{dp_\varphi}{dz} v_z$, and replace $B_r$ with Eq. (9) the following equation can be obtained;

$$\Delta \vec{p}_\varphi = -\frac{q}{2} r \Delta \vec{B}_{(z)} \tag{12}$$

where $r$ is the maximum radius of the beam if we switch to the particle beam concept. According to last equation, the variation of the solenodial magnetic field along the beam axis results in a kick on the transverse plane. This kick could be in the form of focusing or defocusing. $B'_{(z)}$ and $B_r$ are inversely proportional with regard to the Eq.(9) and the beam is focused at the end edge if there is defocusing at the start edge of the solenoid as a result of Eq.(12). This agrees with Figure (5). It can be also seen in Figure (5) that the emittance growth is higher in second solenoid than that of the first solenoid. This is because the beam enters the second solenoid with higher divergence and the higher the divergence when entering the solenoid, the higher the emittance growth when exiting the solenoid [14] as is seen in Figure 6.

4. **Conclusions**

A LEBT channel with two solenoids has been designed for TAC project via PATH MANAGER beam dynamics simulation software. The figures of merit were achieving the input emittance and transmission values obtained in RFQ simulations. For these purposes, we have tuned the magnetic fields of the solenoids using beam dynamics simulation software. We have also examined the simulation results whether they are interpretable analytically. The next study would be comparing solenoids with quadrupoles in the LEBT channel of TAC proton linac.

## 5. Acknowledgements

We thank Turkish State Planning Organization (DPT), under the grants no DPT-2006K120470. We also special thank Alessandra Lombardi from Linac4/CERN and Saleh Sultansoy from TOBB University of Economics and Technology for helps.